\documentclass[aps,pra,preprint,12pt,superscriptaddress]{revtex4-2}

\usepackage{amssymb}
\usepackage{amsmath}
\usepackage{bm}
\usepackage{natbib}
\usepackage{graphics,graphicx}
\usepackage{epsfig}
\usepackage{latexsym}
\usepackage{epic,eepic,graphicx,amssymb,amsmath,indentfirst}
\usepackage{bm}
\usepackage{graphicx}
\usepackage[utf8]{inputenc}
\usepackage[T1]{fontenc}
\usepackage{float}
\usepackage{appendix}

\usepackage{mathrsfs}
\usepackage{calrsfs}

\usepackage{multirow}

\usepackage{color}
\usepackage[colorlinks={true}]{hyperref}
\hypersetup{citecolor={blue}, filecolor={blue}, linkcolor={blue}, urlcolor={blue}}

\begin{document}

\title{\bf Control of the classical dynamics of a particle in the \\ Morse-soft-Coulomb potential}

\author{Gabriel Albertin Amici}
\affiliation{Departamento de F\'isica, Universidade Federal de S\~ao Carlos (UFSCar)\\ S\~ao Carlos, SP 13565-905, Brazil}
\author{José Andrés Guzmán Morán}
\affiliation{Network Science Institute, Northeastern University\\ London, E1W 1LP, United Kingdom}
\author{Emanuel Fernandes de Lima}
\email{eflima@ufscar.br}
\affiliation{Departamento de F\'isica, Universidade Federal de S\~ao Carlos (UFSCar)\\ S\~ao Carlos, SP 13565-905, Brazil}
\date{\today}

\begin{abstract}
We introduce the one-dimensional Morse-soft-Coulomb (MsC) potential consisting of a Morse repulsive barrier smoothly connected with a soft-core Coulomb potential at the origin. This new potential has a single parameter that controls the softness of the repulsive barrier and the well depth. When this softening-depth parameter tends to zero, the MsC potential approaches the Coulomb potential with an infinite repulsive barrier, a known successful model for the hydrogen atom. We investigate the classical chaotic dynamics of the MsC potential subjected to time-dependent external fields, comparing the results with the Coulomb potential. We show that the MsC potential reproduces the dynamics and the ionization probabilities of the Coulomb potential for sufficiently small values of the softening parameter. We also investigate the role of the softening parameter in the phase-space structures, showing that the increasing of its value leads to the increasing of the chaotic sea and consequently to the rise of the ionization probability. Finally, we address the problem of controlling the dynamics of a particle in the MsC potential from the perspective of optimal control theory, which cannot be easily applied in the case of the Coulomb potential due to the singularity at the origin. We analize a particular optimal solution to the problem of transferring a given amount of energy to the system at minimum cost. Our results show that the MsC potential can be a useful simple model for investigating the hydrogen atom.

\end{abstract}
\maketitle


\pagebreak

\section{Introduction}

One-dimensional model potentials are of fundamental importance for Physics since they contribute to a clear understanding of diverse complex atomic and molecular processes such as ionization and dissociation \cite{M_Protopapas_1997,PhysRevA.44.5997,PhysRevA.101.023405,Carrillo-Bernal_2020,PhysRevLett.107.113002}. Standard examples are the Morse potential and the one-dimensional Coulomb potential used, respectively, in the study of dissociation of diatomic molecules and ionization of the hydrogen atom \cite{PhysRevA.79.033416,FORLEVESI2018681,PhysRevE.87.014901,doi:10.1098/rspa.2015.0534,PhysRevA.88.013408,10.1063/1.472058,TMenis_1992}. These models in conjunction with the tools of classical nonlinear dynamics provide important means of gaining physical insight in complicate situations. In particular, the classical approach has been proved to be a powerful tool for analysing the ionization of hydrogen atoms. For instance, the classical driven one-dimensional hydrogen atom yields good agreement with the experimentally determined ionization thresholds \cite{PhysRevLett.55.2231,JENSEN19911,PhysRevLett.89.274101}.

Nonlinear dynamics provides a general description for the escaping of a particle from a potential well, which can be regarded as atomic ionization or molecular dissociation depending on the model potential \cite{PhysRevE.87.014901,CASATI198777,sagdeev1988nonlinear,reichl2013transition,lieberman}. In the absence of external fields, the phase space is composed by a bound region, where the particle undergoes periodic motion, and by an unbound region, where the particle can depart from the potential well. Periodic time-dependent external fields can induce transitions from the bound to the unbound region by means of chaotic routes. For small field strength, some invariant curves remain in phase space, the so-called Kolmogorov–Arnold–Moser (KAM) tori, except in the vicinity of the nonlinear resonances, where resonance islands emerge enclosed by a localized chaotic regions. In this scenario, no escaping occurs, since the motion is still restricted in phase space. However, with increasing field strength, the resonance islands grows and overlap each other leading to the destruction of the KAM tori. The phase-space is then dominated by resonance islands and by a connected chaotic sea, where the particle can diffuse erratically and eventually escapes from the potential well. Although this the the main known escaping mechanism, we mention in passing that there exist also nonchaotic routes which leads to the escaping of the particle through deformed KAM tori \cite{PhysRevE.101.022207}.

When referring to the one-dimensional hydrogen atom model, two distinct Coulomb-like potentials are often considered: one is symmetric around the origin $V(r)\propto -1/|r|$, and the other has an infinite repulsive barrier at the origin $V(r)\propto -1/r$, such that the particle is constrained to move in the positive semi-axis $r\geq 0$. In both cases, the singularity has to be tackled to perform numerical calculations and to this end some techniques based on extended phase space have been developed \cite{JGLeopold_1985,PhysRevA.80.033416}. However, the issue with the singularity becomes more severe, or even prohibitive, when one tries to apply optimal control theory to these problems because the associated Euler-Lagrange equations involves spatial derivatives of the potential function. Since the optimal control of classically chaotic system is an important part of the control of molecular system \cite{PhysRevA.51.923,10.1063/1.4797498,Rabitz2023}, alternative approaches to the classical control of these singular potentials should be developed.

To circumvent the singularity problem in the case of the symmetric Coulomb potential, Eberly and co-workers proposed the soft-Coulomb potential $V(r)\propto -1/\sqrt{r^2+\alpha^2}$, a well-behaved potential with a softening parameter $\alpha$ \cite{nicolaides,INARREA201994,PhysRevE.89.053319,PhysRevA.38.3430,PhysRevLett.64.862,PhysRevA.44.5997,PhysRevLett.112.143006,PhysRevA.48.746}. However, there has been no singularity-free, simple potential to mimic the Coulomb potential with an infinite repulsive barrier.

To fill this gap, we introduce the Morse-soft-Coulomb (MsC) potential consisting of a Morse repulsive barrier smoothly connected with a soft-core Coulomb potential at the origin. This potential is differentiable up to second order while possessing a single parameter that sets the softness of the repulsive barrier and the well depth. The MsC potential approaches the Coulomb potential with an infinite repulsive barrier as the softening-depth parameter tends to zero. We consider the chaotic dynamics of the MsC potential subjected to harmonic driving field and show that the MsC potential reproduces the dynamics and the ionization probabilities of the Coulomb potential for small values of the softening parameter. We analyze the impact of the softening parameter in the phase-space structures, verifying that the increasing of its value leads to the increasing of the chaotic sea and to the rise of the ionization probability. We also address the problem of controlling the dynamics of a particle in the MsC potential using optimal control theory. We consider the problem of transferring a given amount of energy to the system at minimum effort and investigate a particular solution, termed intrinsic optimal solution \cite{PhysRevLett.69.430}.

\section{The Morse-soft-Coulomb potential}

Consider the classical motion of an electron in a new one-dimensional atomic potential, the Morse-soft-Coulomb potential, $V_{ \rm MsC}(r)$. The field-free Hamiltonian describing the particle motion (atomic units are employed throughout the work, unless explicitly stated) is given by,

\begin{equation} \label{H0}
    H_0(r,p)=\frac{p^2}{2}+V_{ \rm MsC}(r).
\end{equation}
The MsC potential consists of a repulsive Morse barrier smoothly jointed to a soft-core Coulomb function and is given by
\begin{equation}
    V_{ \rm MsC}(r)= \begin{cases} 
      -1/\sqrt{r^2+\alpha^2} & r > 0 \\
      \frac{1}{\alpha} \left (e^{-2 r/(\alpha\sqrt{2})}-2e^{- r/(\alpha\sqrt{2})}\right) & r \leq 0, 
      \end{cases}
\end{equation}
where $\alpha$ sets the depth of the potential well, given by $\alpha^{-1}$, as well as the softness of the repulsive barrier. The potential $V_{ \rm MsC}(r)$, its first and second derivatives with respect to the position are continuous. For $\alpha\rightarrow0$, $V_{ \rm MsC}(r)$ tends to the Coulomb potential. In order to allow for comparisons involving the ground level of the hydrogen atom, we will constrain the values of $\alpha$ to be less or equal to $2$.

\begin{figure}[H]
\begin{center}
\includegraphics[width=11cm]{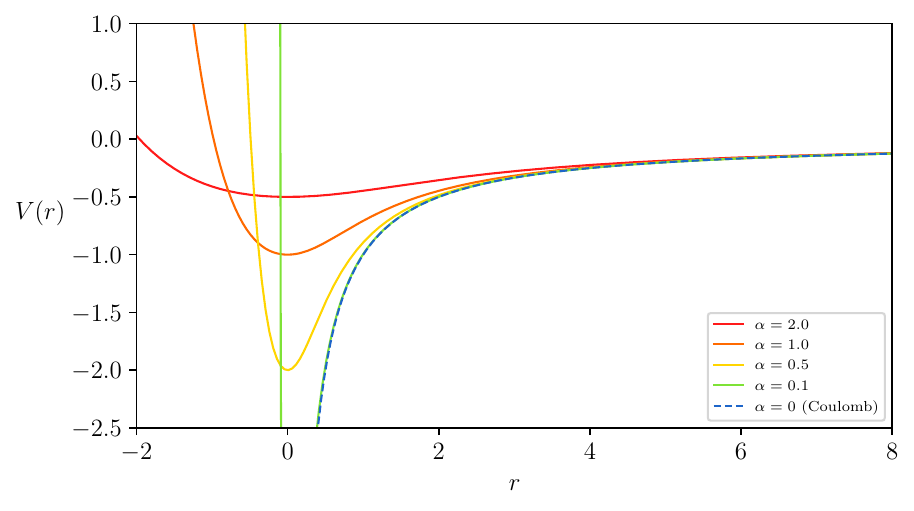}
\end{center}
\caption{\label{MCpot} MsC potential as a function of $r$ for selected values of the $\alpha$ parameter and the Coulomb potential  which corresponds $\alpha=0$ (dashed).}
\end{figure}

Figure~\ref{MCpot} (a) shows the MsC potential for three different values of $\alpha$. For negative energies, the particle describes bound motion (libration), while the particle describes unbound motion for positive energies. In general, it can be noted the good agreement of the MsC potential with the Coulomb potential in the long range, specially for lower values of $\alpha$.

The bound oscillatory motion of the electron can be conveniently analised in terms of the action-angle variables $(J,\theta)$. We have calculated numerically the action defined by,

\begin{equation}\label{action}
    J(E)=\frac{1}{2\pi}\oint p(E,r)dr,
\end{equation}
where $E$ is the energy. From Eq.~(\ref{action}), we interpolate the Hamiltonian $H_0$ for a set of values of the action $J$. Then the frequency is determined by the derivative $\omega(J)=\partial H_0/\partial J$.

 \begin{figure}[ht!]
\begin{center}
\includegraphics[width=10cm]{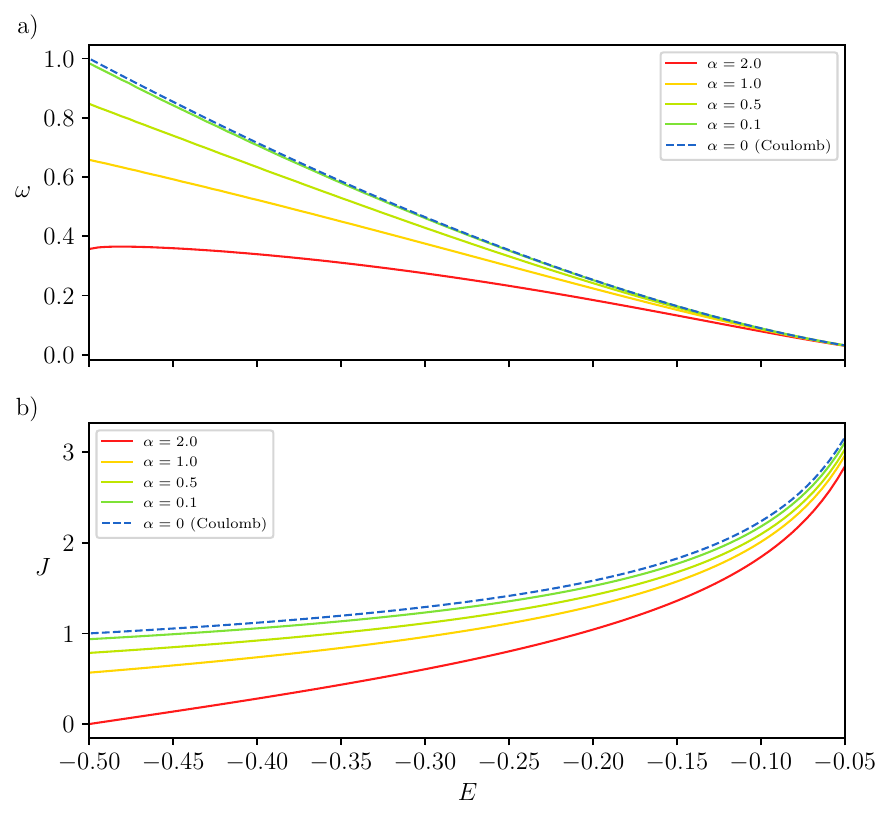}
\end{center}
\caption{\label{MCAction} Relevant quantities calculated for a bound state of the Morse-Coulomb potential compared to Coulomb potential (dashed). a) Angular frequency $\omega$ as a function of the energy $E$. b) Classical action $J$ as a function of the energy.}
\end{figure}
 
 Figure~\ref{MCAction} (a) shows the MsC oscillator frequency $\omega$ as a function of the energy for different values of $\alpha$. The corresponding frequency of the Coulomb potential is also shown in this panel with dotted curve. Panel (b) of Fig.~\ref{MCAction} shows the MsC action as a function of the energy. We note that in both curves the Coulomb behavior is approached as $\alpha \rightarrow0$. It is also evident that lower energies are considerably affected by the softening of the repulsive barrier (the increasing of $\alpha$). We observe that the barrier softening leads to a decreases of the oscillator frequency for a fixed energy: the particle spends a longer time colliding with the softer barrier, meaning that the particle spends more time to reverse its momentum when colliding with the inner barrier. We also note that for lager values of $\alpha$, the changing of the frequency with respect to the energy $|d\omega/dE|$ is smaller.  

\begin{figure}[hb!]
\begin{center}
\includegraphics[width=10cm]{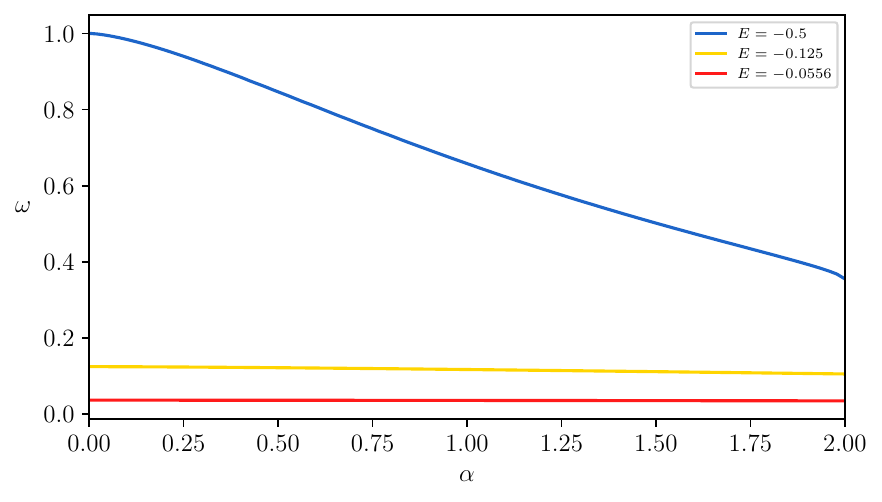}
\end{center}
\caption{\label{Omegalpha} Oscillating frequency as a function of the softening parameter $\alpha$ for the ground energy level $E=-0.5$.}
\end{figure}

 Figure~\ref{Omegalpha} shows the frequency as a function of the softening parameter $\alpha$ considering the ground level energy $E=-0.5$. It can be noted that for $0<\alpha<0.1$ the derivative $|d\omega/d\alpha|$ is relatively small and frequency remains close to the Coulomb frequency, but for increasing $\alpha$ the oscillation frequency decreases considerably, being approximately $0.4$ for $\alpha=2.$

\section{The harmonically driven system}

We consider now the system driven by an harmonic perturbation, which stands for a cw field acting on the electron moving in the MsC potential. To this end, we add to the unperturbed Hamiltonian $H_0$ the interaction Hamiltonian $H_{\rm int}$ given by,

\begin{equation}
    H_{\rm int}(r,t)=rF_0\cos(\Omega t),
\end{equation}
where $F_0$ is the amplitude and $\Omega$ is the frequency of the external field. The system dynamics is now described by the total Hamiltonian $H(t)=H_0+H_{\rm int}$.

We are interested in the chaotic dynamics associated with $H(t)$, particularly in the escaping of trajectories from the potential well, i.e., with the ionization process. As usual to allow for quantum-classical comparison, we can define an ionization probability by distributing a large number of trajectories $N_t$ in a given energy level, which corresponds to an energy level of the hydrogen atom. The initial conditions are distributed uniformly in the angle variable. Then, the classical ionization probability is calculated by counting the number of escaping trajectories $n_e$ by the end of the external perturbation and calculating the fraction,

\begin{equation}
    P_i=\frac{n_e}{N_t},
\end{equation}
We have tested the convergence of $P_i$ with the total number of trajectories and found that $P_i$ will deviate only up to $2\%$ from $N_t=200$.

\begin{figure}[ht!]
\begin{center}
\includegraphics[width=10cm]{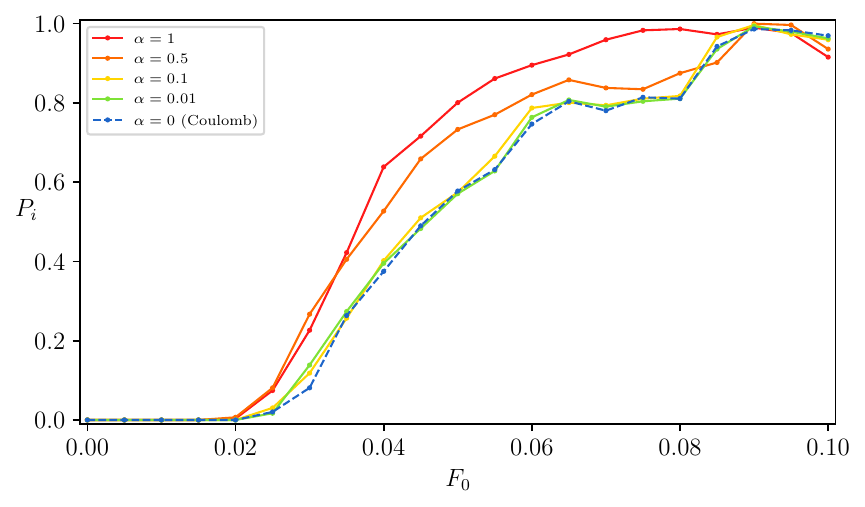}
\end{center}
\caption{\label{Ampli} Ionization probability for several values of $\alpha$ as a function of the amplitude $F_0$ of the external field for $\Omega=1$ and for the initial conditions in the ground energy level $E=-0.5$. The ionization probability for the Coulomb potential $\alpha=0$ is also shown (dashed line).}
\end{figure}

Figure~\ref{Ampli} shows the ionization probability for distinct values of $\alpha$ as a function of the amplitude $F_0$ of the external field for $\Omega=1$ and for the initial conditions distributed in the ground energy level $E=-0.5$. We note that as $\alpha$ is decreased the ionization probability  approaches that of the Coulomb potential, in fact, for $\alpha$ smaller than $0.1$ the ionization probabilities are very close to each other. On the other hand, for the increasing values of $\alpha$ the ionization probability generally also increases. This is an indication of the role of the parameter $\alpha$ in promoting the chaotic ionization as we will show latter.

\begin{figure}[hb!]
\begin{center}
\includegraphics[width=10cm]{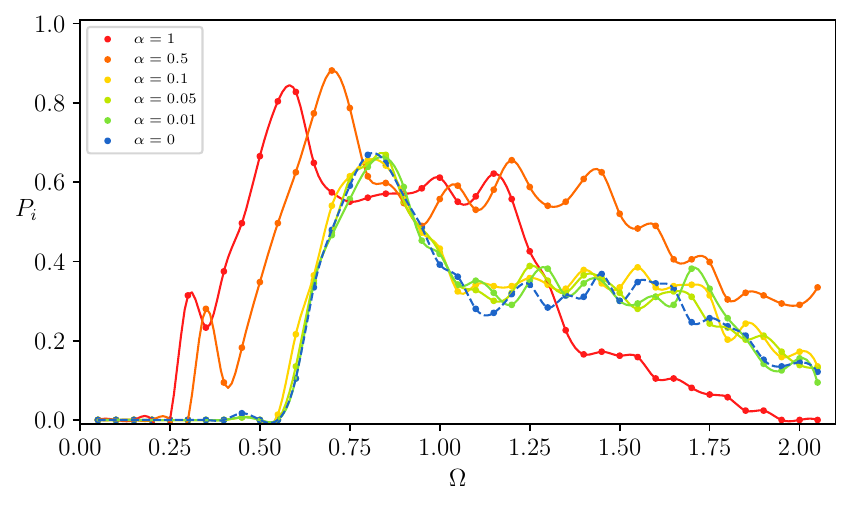}
\end{center}
\caption{\label{Freq} Ionization probability for several values of $\alpha$ as a function of the frequency $\Omega$ of the external field for $F_0=0.04$ and for the initial conditions in the ground energy level $E=-0.5$. The ionization probability for the Coulomb potential $\alpha=0$ is also shown (dashed line).}
\end{figure}

Figure~\ref{Freq} shows the ionization probability for distinct values of $\alpha$ as a function of the frequency $\Omega$ of the external field for $F_0=0.04$ and for the initial conditions distributed in the ground energy level $E=-0.5$. We note that as $\alpha$ is decreased the ionization probability  approaches that of the Coulomb potential. On the other hand, for increasing values of $\alpha$ the main peak of the ionization probability is red shifted. This behaviour can be attributed to the frequency of the MsC potential, that also decreases for increasing $\alpha$, as shown in Fig.\ref{Omegalpha}.

\begin{figure}[ht!]
\begin{center}
\includegraphics[width=10cm]{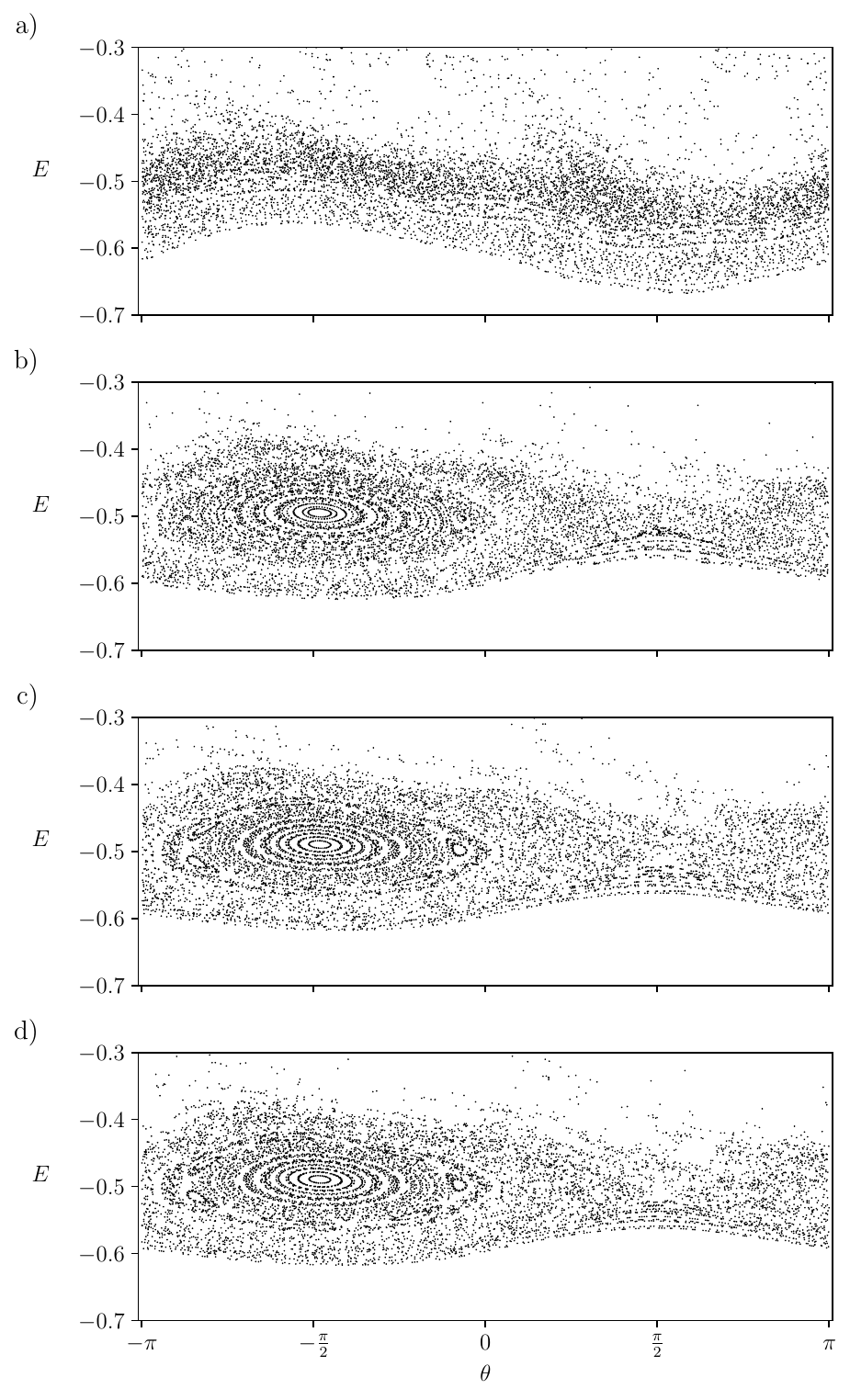}
\end{center}
\caption{\label{poincare} Poincaré surface of sections for distinct values of the softening parameter: (a) $\alpha=1.$, (b) $\alpha=0.1$, (c) $\alpha=0.01$ and (d) $\alpha=0$. In all panels, the field amplitude is $F_0=0.025$ and the frequency is $\Omega=1$.}
\end{figure}

The surface of section (stroboscopic map) energy versus angle for several values of $\alpha$ are shown in Fig.~\ref{poincare}. They were build by distributing $180$ initial conditions over a small energy interval around $E=-0.5$ and marking the point of the trajectory in the energy-angle plane for each period of the external field, i.e., for $t=n\times 2\pi/\Omega$, for positive integer $n$, $n=0,1,\ldots$. In all panels the field amplitude is $F_0=0.025$ and the frequency is $\Omega=1$. In panel (a), $\alpha=1$ and any structure are hardly seen in the map, composed essentially by a chaotic sea. In panel (b), $\alpha=0.1$ and the large $1:1$ resonance island becomes evident as some KAM tori for lower energies. Panels (c) and (d), where $\alpha=0.01$ and $\alpha=0$, respectively, are almost indistinguishable. Apart from the structures seen in panel (b), some secondary resonance island are also observed. From these panels, we note that as $\alpha$ is decreased the dynamics becomes very close to the dynamics of the Coulomb potential with a large dominant resonant island, whereas as $\alpha$ is increased the chaotic dynamics dominates the phase-space. We thus conclude that the increasing of the ionization probabilities for increasing $\alpha$ observed in Figs.~\ref{Ampli} and \ref{Freq} can be attributed to the disappearing (collapse) of the primary resonance island and the formation of a large chaotic sea, where the trajectories can diffuse up to the ionization threshold.

\pagebreak

\pagebreak

\section{Optimal energy transfer}\label{OET}

As an application of the MsC potential, we consider the optimal control problem of performing a fixed energy transfer from an electromagnetic field to an atomic system in a specified initial state under minimum cost\cite{PhysRevLett.69.430}. Thus, it is desired to find a control function $u(t)$ (an external electromagnetic field) acting from $t=0$ to $t=t_f$ that preforms a given energy transfer to the system (an electron in a MsC potential) in such a way that minimizes the cost functional given by,

\begin{align}
    J\{u(t)\}=\int_0^{t_f}u(t)^2dt,
\end{align}
which accounts for the field fluence and represents a measure of the effort of the control field. The system dynamics is governed by the total Hamiltonian,

\begin{equation}\label{tdH}
    H(t)=H_0(r,p)+ ru(t)   ,
\end{equation}
and consequently it is constrained to obey the equations of motion,

\begin{align}\label{eqm1}
  &  \dot{r}(t)=p(t), \\
  &   \dot{p}(t)=-\frac{d }{dr}V_{\rm MsC}- u(t), \label{eqm2}
\end{align}
with the initial conditions $r(0)=r_0$ and $p(0)=p_0$.

Finally, a certain fixed amount of energy has to be transferred to the system during the action of the control field, which is represented by an extra constraint,

\begin{equation}\label{isop}
   \int_0^{t_f}-u(t) p(t)dt=C,
\end{equation}
with $C$ being a constant.

We can form the so-called control Hamiltonian $\mathcal{H}$ (not to be confused with the Hamiltonian in Eq.(\ref{tdH})) for the above control problem \cite{kirk},

\begin{equation}\label{cH}
    \mathcal{H}=u(t)^2+\lambda_1(t)p(t)-\lambda_2(t)\left[\frac{d }{dr}V_{\rm MsC}+ u(t)\right]-e\lambda_3 u(t)p(t),
\end{equation}
where $\lambda_1(t)$ and $\lambda_2(t)$ are the time-dependent Lagrange multipliers, or the costates, associated with $r(t)$ and $p(t)$, respectively, while $\lambda_3$ is a non-zero constant Lagrange multiplier associated with the integral constraint, Eq.(\ref{isop}).

From the control Hamiltonian, we can derive the necessary conditions for the optimal fields. Apart from the equations of motion (\ref{eqm1}) and (\ref{eqm2}), we obtain the costate equations,
\begin{align}\label{cseq1}
  &  \dot{\lambda}_1(t)=-\frac{\partial \mathcal{H}}{\partial r}= \lambda_2(t)\frac{d^2}{dr^2} V_{\rm MsC}, \\
  &   \dot{\lambda}_2(t)=-\frac{\partial \mathcal{H}}{\partial p}=-\lambda_1(t)+\lambda_3 u(t), \label{cseq2}
\end{align}
along with the final-time conditions $\lambda_1(t_f)=\lambda_2(t_t)=0$. Furthermore, we obtain the equation for the control field,

\begin{equation}\label{grad}
    \frac{\partial \mathcal{H}}{\partial u}=2u(t)-\lambda_2(t)-\lambda_3p(t)=0.
\end{equation}

Differentiating twice Eq.(\ref{grad}) with respect to time and using the equations of motion and the costate equations, we obtain the following second-order differential equation for the control field,

\begin{equation}\label{SOEfield}
    \ddot{u}(t)+u(t)\frac{d^2 }{dr^2}V_{\rm MsC}=0,
\end{equation}
accompanied by the final-time conditions,

\begin{align}\label{ftc1}
  &  u(t_f)= \frac{\lambda_3}{2}p(t_f), \\
  &  \dot{u}(t_f)= -\frac{\lambda_3}{2}\frac{d }{dr}V_{\rm MsC}(r(t_f)),\label{ftc2}
\end{align}
where the value of $\lambda_3$ sets the constant $C$. Equation (\ref{SOEfield}) and the equations of motion (\ref{eqm1}) and (\ref{eqm2}) along with the initial conditions $(r_0,p_0)$ and final-time conditions  (\ref{ftc1}) and (\ref{ftc2}) stand for the necessary conditions to be met by an optimal solution. Usually, the solution for these equations are sought numerically. We note that the second derivative in Eq.~(\ref{SOEfield}) may turn such task impractical for a singular potential, as for the Coulomb potential. An alternative would then be to replace the Coulomb potential by the MsC potential with appropriate value of the softening parameter. 

Here, instead of solving the above two-point boundary value problem, we focus on a particular form of solution of Eq.(\ref{SOEfield}) given by \cite{PhysRevLett.69.430},

\begin{equation}\label{intrinsic}
    u^*(t)=-\frac{2p(t)}{(t+\tilde{t})},
\end{equation}
where $\tilde{t}$ is a constant. The final-time conditions for this solution yield $\lambda_3=4/(t_f+\tilde{t})^{-1}$ and $u^*(t_f)=0$.

 \begin{figure}[ht!]
\begin{center}
\includegraphics[width=15cm]{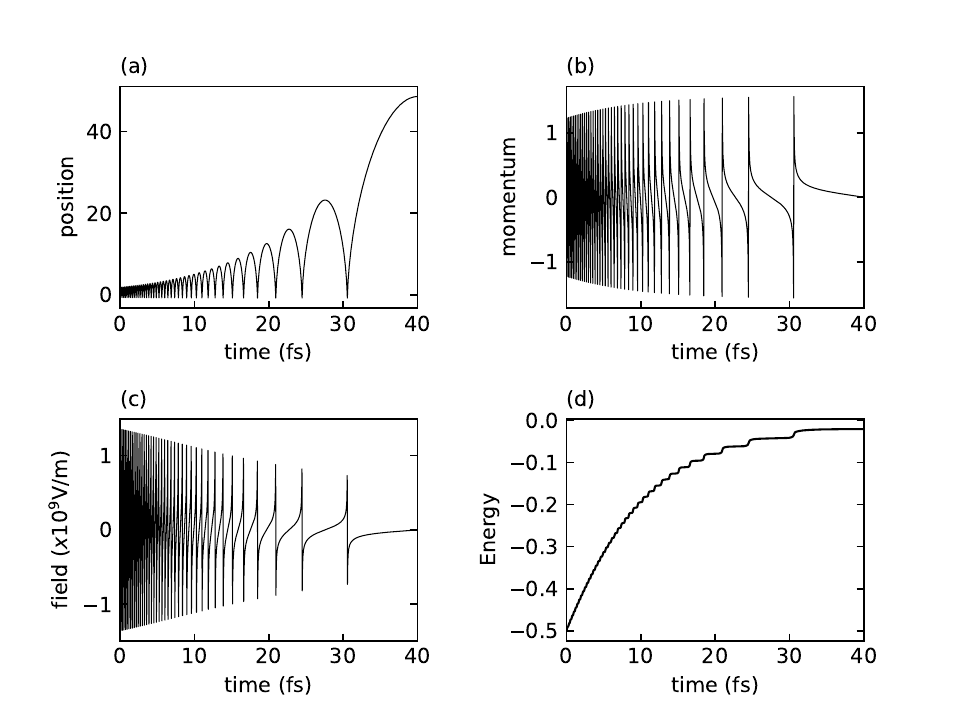}
\end{center}
\caption{\label{Fig:Opt1} Action of the intrinsic optimal field on the system dynamics. Softening parameter $\alpha=0.8$, $t_f=40$fs, $\tilde{t}=22.47$fs, initial energy $E_i=-0.5$, $r_0=0.9r_{ret}$, where $r_{ret}$ is the positive point of return of the MsC potential for $E_i$. The initial momentum is calculated by $p_0=-\sqrt{2(E_i-V(r_0))}$. (a) and (b) position and momentum, respectively, as a function of time. (c) Optimal control field. (d) Energy of the unperturbed system $H_0(r,p)$. }
\end{figure}

Since the choice of $\tilde{t}$ specifies the values of $u(0)$, we can simply solve the initial-value problem given by the equations of motion (\ref{eqm1}) and (\ref{eqm2}) with $u(t)$ given by Eq.~(\ref{intrinsic}). Therefore, $\tilde{t}$ can be adjusted to fulfill the necessary final-time condition $u^*(t_f)=0$.

Figure~\ref{Fig:Opt1} illustrates the results obtained for an optimal control field, where we have set the softening parameter $\alpha=0.8$, the final time $t_f=40$fs and $\tilde{t}=22.47$fs. The initial condition was chosen with energy $E_i=-0.5$ and position $r_0=0.9r_{ret}$ , where $r_{ret}$ is the positive point of return of the MsC potential for $E_i$. The initial momentum was set to $p_0=-\sqrt{2(E_i-V(r_0))}$. Panels (a) and (b) show respectively the position and momentum of the particle. We observe librational motion of increasing amplitude, the abrupt changes in position and momentum corresponding to the collisions with hard repulsive barrier. The control field shown in panel (c) presents a negative chirping feature, which can be related to the decreasing of the oscillator frequency as the energy is increased (see panel (a) of Fig.\ref{MCAction}). Panel (d) shows the energy stored in the field-free system, $H_0(r,p)$. We note that the transferred energy was $C=0.48$ and that the particle almost reach the ionization threshold. Interestingly, the energy increases in a stair-like way. Comparing panel (d) to panel (c) we note that each rung corresponds to the collision with the hard repulsive barrier. Thus, an optimal way to transfer energy is to push the particle against the barrier and rapidly to pull out from the barrier, following the changing of the particle momentum.

\begin{figure}[hb!]
\begin{center}
\includegraphics[width=15cm]{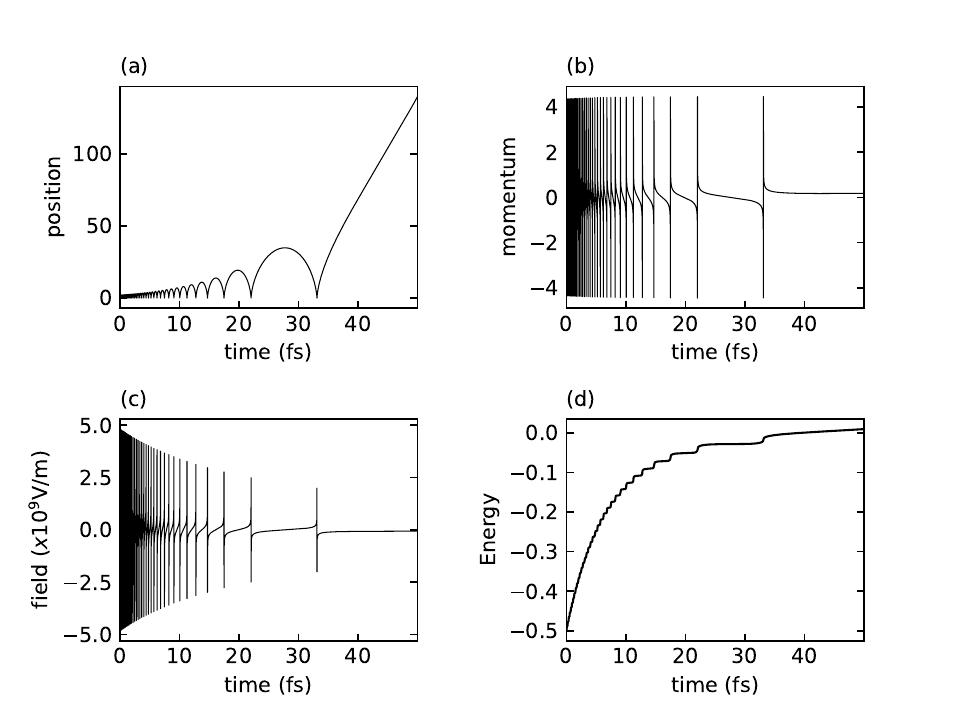}
\end{center}
\caption{\label{Fig:Opt2} Action of the intrinsic optimal field on the system dynamics. Softening parameter $\alpha=0.1$, $t_f=50$fs, $\tilde{t}=22.47$fs, initial energy $E_i=-0.5$, $r_0=0.9r_{ret}$, where $r_{ret}$ is the positive point of return of the MsC potential for $E_i$. The initial momentum is calculated by $p_0=-\sqrt{2(E_i-V(r_0))}$. (a) and (b) position and momentum, respectively, as a function of time. (c) Optimal control field. (d) Energy of the unperturbed system $H_0(r,p)$. }
\end{figure}

In Fig.~\ref{Fig:Opt2}, we consider a solution of Eq.(\ref{SOEfield}) which is not optimal for it does not fulfill the end-time condition $u(t_f)=0$. Nevertheless, the field leads to ionization, as can be seen in panel (d), by the same mechanism discussed above. It is interesting to note that once the particle escapes from the potential well with sufficient kinetic energy it may not return to the potential, and the momentum will not change its sign. Thus, in these ionization cases, the control field given in Eq.(\ref{SOEfield}) cannot be optimal.

\section{Conclusions}

This work presented the smooth and singularity -free Morse-soft-Coulomb potential model, which has a single parameter that control the softening of the repulsive barrier and the well depth. The MsC tends to the Coulomb potential with an infinite repulsive barrier as the softening parameter tends to zero. We investigated this model classically using nonlinear dynamics tools. We have obtained good agreement between the results of ionization probabilities between the MsC and the Coulomb potential for $\alpha\leq 0.1$. We also verified the effect of increasing the softening parameter, which leads to an increase of the ionization probability. Analysing the stroboscopic maps, we observed that the rise of the softening parameter led to a destruction of the resonance island and an increasing of the chaotic sea. This fact can be attributed to the fact that increasing $\alpha$ leads to 
 a overall decreasing of the oscillator frequency and also leads to the decreasing of the ratio in which the frequency changes with energy. Thus, primary resonance conditions with the external field will become closer in phase space, favouring resonance overlapping and the consequent rise of ionization. Finally, we have addressed the optimal control problem of transfer a certain amount of energy to the system at minimum cost of the external control field. We have considered a particular solution termed intrinsic optimal solution. For these solutions the transfer mechanism consists of essentially of pushing the particle against the inner repulsive barrier and rapidly pulling it out from the barrier  accompanying the particle momentum as it collides. We expect that the MsC potential can be investigated further and also may serve as a model beyond the hydrogen atom. The investigation of the quantum version of the potential is being carried out and will be presented in next works. 

\section*{Acknowledgments}\label{sec7}
The authors acknowledge support from the Brazilian agency S\~ao Paulo Research Foundation, FAPESP (grants 23/04930-4, 23/00690-9, 19/15679-5 and  14/23648-9).


\begin{thebibliography}{34}%
\makeatletter
\providecommand \@ifxundefined [1]{%
 \@ifx{#1\undefined}
}%
\providecommand \@ifnum [1]{%
 \ifnum #1\expandafter \@firstoftwo
 \else \expandafter \@secondoftwo
 \fi
}%
\providecommand \@ifx [1]{%
 \ifx #1\expandafter \@firstoftwo
 \else \expandafter \@secondoftwo
 \fi
}%
\providecommand \natexlab [1]{#1}%
\providecommand \enquote  [1]{``#1''}%
\providecommand \bibnamefont  [1]{#1}%
\providecommand \bibfnamefont [1]{#1}%
\providecommand \citenamefont [1]{#1}%
\providecommand \href@noop [0]{\@secondoftwo}%
\providecommand \href [0]{\begingroup \@sanitize@url \@href}%
\providecommand \@href[1]{\@@startlink{#1}\@@href}%
\providecommand \@@href[1]{\endgroup#1\@@endlink}%
\providecommand \@sanitize@url [0]{\catcode `\\12\catcode `\$12\catcode
  `\&12\catcode `\#12\catcode `\^12\catcode `\_12\catcode `\%12\relax}%
\providecommand \@@startlink[1]{}%
\providecommand \@@endlink[0]{}%
\providecommand \url  [0]{\begingroup\@sanitize@url \@url }%
\providecommand \@url [1]{\endgroup\@href {#1}{\urlprefix }}%
\providecommand \urlprefix  [0]{URL }%
\providecommand \Eprint [0]{\href }%
\providecommand \doibase [0]{https://doi.org/}%
\providecommand \selectlanguage [0]{\@gobble}%
\providecommand \bibinfo  [0]{\@secondoftwo}%
\providecommand \bibfield  [0]{\@secondoftwo}%
\providecommand \translation [1]{[#1]}%
\providecommand \BibitemOpen [0]{}%
\providecommand \bibitemStop [0]{}%
\providecommand \bibitemNoStop [0]{.\EOS\space}%
\providecommand \EOS [0]{\spacefactor3000\relax}%
\providecommand \BibitemShut  [1]{\csname bibitem#1\endcsname}%
\let\auto@bib@innerbib\@empty
\bibitem [{\citenamefont {Protopapas}\ \emph {et~al.}(1997)\citenamefont
  {Protopapas}, \citenamefont {Keitel},\ and\ \citenamefont
  {Knight}}]{M_Protopapas_1997}%
  \BibitemOpen
  \bibfield  {author} {\bibinfo {author} {\bibfnamefont {M.}~\bibnamefont
  {Protopapas}}, \bibinfo {author} {\bibfnamefont {C.~H.}\ \bibnamefont
  {Keitel}},\ and\ \bibinfo {author} {\bibfnamefont {P.~L.}\ \bibnamefont
  {Knight}},\ }\bibfield  {title} {\bibinfo {title} {Atomic physics with
  super-high intensity lasers},\ }\href
  {https://doi.org/10.1088/0034-4885/60/4/001} {\bibfield  {journal} {\bibinfo
  {journal} {Reports on Progress in Physics}\ }\textbf {\bibinfo {volume}
  {60}},\ \bibinfo {pages} {389} (\bibinfo {year} {1997})}\BibitemShut
  {NoStop}%
\bibitem [{\citenamefont {Su}\ and\ \citenamefont
  {Eberly}(1991)}]{PhysRevA.44.5997}%
  \BibitemOpen
  \bibfield  {author} {\bibinfo {author} {\bibfnamefont {Q.}~\bibnamefont
  {Su}}\ and\ \bibinfo {author} {\bibfnamefont {J.~H.}\ \bibnamefont
  {Eberly}},\ }\bibfield  {title} {\bibinfo {title} {Model atom for multiphoton
  physics},\ }\href {https://doi.org/10.1103/PhysRevA.44.5997} {\bibfield
  {journal} {\bibinfo  {journal} {Phys. Rev. A}\ }\textbf {\bibinfo {volume}
  {44}},\ \bibinfo {pages} {5997} (\bibinfo {year} {1991})}\BibitemShut
  {NoStop}%
\bibitem [{\citenamefont {Majorosi}\ \emph {et~al.}(2020)\citenamefont
  {Majorosi}, \citenamefont {Benedict}, \citenamefont {Bog\'ar}, \citenamefont
  {Paragi},\ and\ \citenamefont {Czirj\'ak}}]{PhysRevA.101.023405}%
  \BibitemOpen
  \bibfield  {author} {\bibinfo {author} {\bibfnamefont {S.}~\bibnamefont
  {Majorosi}}, \bibinfo {author} {\bibfnamefont {M.~G.}\ \bibnamefont
  {Benedict}}, \bibinfo {author} {\bibfnamefont {F.}~\bibnamefont {Bog\'ar}},
  \bibinfo {author} {\bibfnamefont {G.}~\bibnamefont {Paragi}},\ and\ \bibinfo
  {author} {\bibfnamefont {A.}~\bibnamefont {Czirj\'ak}},\ }\bibfield  {title}
  {\bibinfo {title} {Density-based one-dimensional model potentials for
  strong-field simulations in {He}, ${{\mathrm{H}}}_{2}{}^{+}$, and
  ${{\mathrm{H}}}_{2}$},\ }\href {https://doi.org/10.1103/PhysRevA.101.023405}
  {\bibfield  {journal} {\bibinfo  {journal} {Phys. Rev. A}\ }\textbf {\bibinfo
  {volume} {101}},\ \bibinfo {pages} {023405} (\bibinfo {year}
  {2020})}\BibitemShut {NoStop}%
\bibitem [{\citenamefont {Carrillo-Bernal}\ \emph {et~al.}(2020)\citenamefont
  {Carrillo-Bernal}, \citenamefont {y~Romero}, \citenamefont {Núñez-Yépez},
  \citenamefont {Salas-Brito},\ and\ \citenamefont
  {Solis}}]{Carrillo-Bernal_2020}%
  \BibitemOpen
  \bibfield  {author} {\bibinfo {author} {\bibfnamefont {M.~A.}\ \bibnamefont
  {Carrillo-Bernal}}, \bibinfo {author} {\bibfnamefont {R.~P.~M.}\ \bibnamefont
  {y~Romero}}, \bibinfo {author} {\bibfnamefont {H.~N.}\ \bibnamefont
  {Núñez-Yépez}}, \bibinfo {author} {\bibfnamefont {A.~L.}\ \bibnamefont
  {Salas-Brito}},\ and\ \bibinfo {author} {\bibfnamefont {D.~A.}\ \bibnamefont
  {Solis}},\ }\bibfield  {title} {\bibinfo {title} {Classical and quantum space
  splitting: the one-dimensional hydrogen atom},\ }\href
  {https://doi.org/10.1088/1361-6404/aba78e} {\bibfield  {journal} {\bibinfo
  {journal} {European Journal of Physics}\ }\textbf {\bibinfo {volume} {41}},\
  \bibinfo {pages} {065405} (\bibinfo {year} {2020})}\BibitemShut {NoStop}%
\bibitem [{\citenamefont {Burke}\ \emph {et~al.}(2011)\citenamefont {Burke},
  \citenamefont {Mitchell}, \citenamefont {Wyker}, \citenamefont {Ye},\ and\
  \citenamefont {Dunning}}]{PhysRevLett.107.113002}%
  \BibitemOpen
  \bibfield  {author} {\bibinfo {author} {\bibfnamefont {K.}~\bibnamefont
  {Burke}}, \bibinfo {author} {\bibfnamefont {K.~A.}\ \bibnamefont {Mitchell}},
  \bibinfo {author} {\bibfnamefont {B.}~\bibnamefont {Wyker}}, \bibinfo
  {author} {\bibfnamefont {S.}~\bibnamefont {Ye}},\ and\ \bibinfo {author}
  {\bibfnamefont {F.~B.}\ \bibnamefont {Dunning}},\ }\bibfield  {title}
  {\bibinfo {title} {Demonstration of turnstiles as a chaotic ionization
  mechanism in rydberg atoms},\ }\href
  {https://doi.org/10.1103/PhysRevLett.107.113002} {\bibfield  {journal}
  {\bibinfo  {journal} {Phys. Rev. Lett.}\ }\textbf {\bibinfo {volume} {107}},\
  \bibinfo {pages} {113002} (\bibinfo {year} {2011})}\BibitemShut {NoStop}%
\bibitem [{\citenamefont {Sethi}\ and\ \citenamefont
  {Keshavamurthy}(2009)}]{PhysRevA.79.033416}%
  \BibitemOpen
  \bibfield  {author} {\bibinfo {author} {\bibfnamefont {A.}~\bibnamefont
  {Sethi}}\ and\ \bibinfo {author} {\bibfnamefont {S.}~\bibnamefont
  {Keshavamurthy}},\ }\bibfield  {title} {\bibinfo {title} {Local phase space
  control and interplay of classical and quantum effects in dissociation of a
  driven morse oscillator},\ }\href
  {https://doi.org/10.1103/PhysRevA.79.033416} {\bibfield  {journal} {\bibinfo
  {journal} {Phys. Rev. A}\ }\textbf {\bibinfo {volume} {79}},\ \bibinfo
  {pages} {033416} (\bibinfo {year} {2009})}\BibitemShut {NoStop}%
\bibitem [{\citenamefont {Forlevesi}\ \emph {et~al.}(2018)\citenamefont
  {Forlevesi}, \citenamefont {{Egydio de Carvalho}},\ and\ \citenamefont {{de
  Lima}}}]{FORLEVESI2018681}%
  \BibitemOpen
  \bibfield  {author} {\bibinfo {author} {\bibfnamefont {M.}~\bibnamefont
  {Forlevesi}}, \bibinfo {author} {\bibfnamefont {R.}~\bibnamefont {{Egydio de
  Carvalho}}},\ and\ \bibinfo {author} {\bibfnamefont {E.}~\bibnamefont {{de
  Lima}}},\ }\bibfield  {title} {\bibinfo {title} {A tunable mechanism to
  control photo-dissociation with invariant tori with variable energies},\
  }\href {https://doi.org/https://doi.org/10.1016/j.physa.2017.08.131}
  {\bibfield  {journal} {\bibinfo  {journal} {Physica A: Statistical Mechanics
  and its Applications}\ }\textbf {\bibinfo {volume} {490}},\ \bibinfo {pages}
  {681} (\bibinfo {year} {2018})}\BibitemShut {NoStop}%
\bibitem [{\citenamefont {de~Lima}\ \emph {et~al.}(2013)\citenamefont
  {de~Lima}, \citenamefont {Ramos},\ and\ \citenamefont
  {de~Carvalho}}]{PhysRevE.87.014901}%
  \BibitemOpen
  \bibfield  {author} {\bibinfo {author} {\bibfnamefont {E.~F.}\ \bibnamefont
  {de~Lima}}, \bibinfo {author} {\bibfnamefont {T.~N.}\ \bibnamefont {Ramos}},\
  and\ \bibinfo {author} {\bibfnamefont {R.~E.}\ \bibnamefont {de~Carvalho}},\
  }\bibfield  {title} {\bibinfo {title} {Role of the range of the dipole
  function in the classical dynamics of molecular dissociation},\ }\href
  {https://doi.org/10.1103/PhysRevE.87.014901} {\bibfield  {journal} {\bibinfo
  {journal} {Phys. Rev. E}\ }\textbf {\bibinfo {volume} {87}},\ \bibinfo
  {pages} {014901} (\bibinfo {year} {2013})}\BibitemShut {NoStop}%
\bibitem [{\citenamefont {Loudon}(2016)}]{doi:10.1098/rspa.2015.0534}%
  \BibitemOpen
  \bibfield  {author} {\bibinfo {author} {\bibfnamefont {R.}~\bibnamefont
  {Loudon}},\ }\bibfield  {title} {\bibinfo {title} {One-dimensional hydrogen
  atom},\ }\href {https://doi.org/10.1098/rspa.2015.0534} {\bibfield  {journal}
  {\bibinfo  {journal} {Proceedings of the Royal Society A: Mathematical,
  Physical and Engineering Sciences}\ }\textbf {\bibinfo {volume} {472}},\
  \bibinfo {pages} {20150534} (\bibinfo {year} {2016})}\BibitemShut {NoStop}%
\bibitem [{\citenamefont {Burke}\ \emph {et~al.}(2013)\citenamefont {Burke},
  \citenamefont {Mitchell}, \citenamefont {Ye},\ and\ \citenamefont
  {Dunning}}]{PhysRevA.88.013408}%
  \BibitemOpen
  \bibfield  {author} {\bibinfo {author} {\bibfnamefont {K.}~\bibnamefont
  {Burke}}, \bibinfo {author} {\bibfnamefont {K.~A.}\ \bibnamefont {Mitchell}},
  \bibinfo {author} {\bibfnamefont {S.}~\bibnamefont {Ye}},\ and\ \bibinfo
  {author} {\bibfnamefont {F.~B.}\ \bibnamefont {Dunning}},\ }\bibfield
  {title} {\bibinfo {title} {Chaotic ionization of a stationary electron state
  via a phase space turnstile},\ }\href
  {https://doi.org/10.1103/PhysRevA.88.013408} {\bibfield  {journal} {\bibinfo
  {journal} {Phys. Rev. A}\ }\textbf {\bibinfo {volume} {88}},\ \bibinfo
  {pages} {013408} (\bibinfo {year} {2013})}\BibitemShut {NoStop}%
\bibitem [{\citenamefont {Korolkov}\ \emph {et~al.}(1996)\citenamefont
  {Korolkov}, \citenamefont {Paramonov},\ and\ \citenamefont
  {Schmidt}}]{10.1063/1.472058}%
  \BibitemOpen
  \bibfield  {author} {\bibinfo {author} {\bibfnamefont {M.~V.}\ \bibnamefont
  {Korolkov}}, \bibinfo {author} {\bibfnamefont {G.~K.}\ \bibnamefont
  {Paramonov}},\ and\ \bibinfo {author} {\bibfnamefont {B.}~\bibnamefont
  {Schmidt}},\ }\bibfield  {title} {\bibinfo {title} {{State‐selective
  control for vibrational excitation and dissociation of diatomic molecules
  with shaped ultrashort infrared laser pulses}},\ }\href
  {https://doi.org/10.1063/1.472058} {\bibfield  {journal} {\bibinfo  {journal}
  {The Journal of Chemical Physics}\ }\textbf {\bibinfo {volume} {105}},\
  \bibinfo {pages} {1862} (\bibinfo {year} {1996})}\BibitemShut {NoStop}%
\bibitem [{\citenamefont {Menis}\ \emph {et~al.}(1992)\citenamefont {Menis},
  \citenamefont {Taieb}, \citenamefont {Veniard},\ and\ \citenamefont
  {Maquet}}]{TMenis_1992}%
  \BibitemOpen
  \bibfield  {author} {\bibinfo {author} {\bibfnamefont {T.}~\bibnamefont
  {Menis}}, \bibinfo {author} {\bibfnamefont {R.}~\bibnamefont {Taieb}},
  \bibinfo {author} {\bibfnamefont {V.}~\bibnamefont {Veniard}},\ and\ \bibinfo
  {author} {\bibfnamefont {A.}~\bibnamefont {Maquet}},\ }\bibfield  {title}
  {\bibinfo {title} {Stabilization of atoms in superintense laser fields: role
  of the coulomb singularity},\ }\href
  {https://doi.org/10.1088/0953-4075/25/11/001} {\bibfield  {journal} {\bibinfo
   {journal} {Journal of Physics B: Atomic, Molecular and Optical Physics}\
  }\textbf {\bibinfo {volume} {25}},\ \bibinfo {pages} {L263} (\bibinfo {year}
  {1992})}\BibitemShut {NoStop}%
\bibitem [{\citenamefont {van Leeuwen}\ \emph {et~al.}(1985)\citenamefont {van
  Leeuwen}, \citenamefont {Oppen}, \citenamefont {Renwick}, \citenamefont
  {Bowlin}, \citenamefont {Koch}, \citenamefont {Jensen}, \citenamefont {Rath},
  \citenamefont {Richards},\ and\ \citenamefont
  {Leopold}}]{PhysRevLett.55.2231}%
  \BibitemOpen
  \bibfield  {author} {\bibinfo {author} {\bibfnamefont {K.~A.~H.}\
  \bibnamefont {van Leeuwen}}, \bibinfo {author} {\bibfnamefont {G.~v.}\
  \bibnamefont {Oppen}}, \bibinfo {author} {\bibfnamefont {S.}~\bibnamefont
  {Renwick}}, \bibinfo {author} {\bibfnamefont {J.~B.}\ \bibnamefont {Bowlin}},
  \bibinfo {author} {\bibfnamefont {P.~M.}\ \bibnamefont {Koch}}, \bibinfo
  {author} {\bibfnamefont {R.~V.}\ \bibnamefont {Jensen}}, \bibinfo {author}
  {\bibfnamefont {O.}~\bibnamefont {Rath}}, \bibinfo {author} {\bibfnamefont
  {D.}~\bibnamefont {Richards}},\ and\ \bibinfo {author} {\bibfnamefont
  {J.~G.}\ \bibnamefont {Leopold}},\ }\bibfield  {title} {\bibinfo {title}
  {Microwave ionization of hydrogen atoms: Experiment versus classical
  dynamics},\ }\href {https://doi.org/10.1103/PhysRevLett.55.2231} {\bibfield
  {journal} {\bibinfo  {journal} {Phys. Rev. Lett.}\ }\textbf {\bibinfo
  {volume} {55}},\ \bibinfo {pages} {2231} (\bibinfo {year}
  {1985})}\BibitemShut {NoStop}%
\bibitem [{\citenamefont {Jensen}\ \emph {et~al.}(1991)\citenamefont {Jensen},
  \citenamefont {Susskind},\ and\ \citenamefont {Sanders}}]{JENSEN19911}%
  \BibitemOpen
  \bibfield  {author} {\bibinfo {author} {\bibfnamefont {R.}~\bibnamefont
  {Jensen}}, \bibinfo {author} {\bibfnamefont {S.}~\bibnamefont {Susskind}},\
  and\ \bibinfo {author} {\bibfnamefont {M.}~\bibnamefont {Sanders}},\
  }\bibfield  {title} {\bibinfo {title} {Chaotic ionization of highly excited
  hydrogen atoms: Comparison of classical and quantum theory with experiment},\
  }\href {https://doi.org/https://doi.org/10.1016/0370-1573(91)90113-Z}
  {\bibfield  {journal} {\bibinfo  {journal} {Physics Reports}\ }\textbf
  {\bibinfo {volume} {201}},\ \bibinfo {pages} {1} (\bibinfo {year}
  {1991})}\BibitemShut {NoStop}%
\bibitem [{\citenamefont {Sirko}\ and\ \citenamefont
  {Koch}(2002)}]{PhysRevLett.89.274101}%
  \BibitemOpen
  \bibfield  {author} {\bibinfo {author} {\bibfnamefont {L.}~\bibnamefont
  {Sirko}}\ and\ \bibinfo {author} {\bibfnamefont {P.~M.}\ \bibnamefont
  {Koch}},\ }\bibfield  {title} {\bibinfo {title} {Control of common resonances
  in bichromatically driven hydrogen atoms},\ }\href
  {https://doi.org/10.1103/PhysRevLett.89.274101} {\bibfield  {journal}
  {\bibinfo  {journal} {Phys. Rev. Lett.}\ }\textbf {\bibinfo {volume} {89}},\
  \bibinfo {pages} {274101} (\bibinfo {year} {2002})}\BibitemShut {NoStop}%
\bibitem [{\citenamefont {Casati}\ \emph {et~al.}(1987)\citenamefont {Casati},
  \citenamefont {Chirikov}, \citenamefont {Shepelyansky},\ and\ \citenamefont
  {Guarneri}}]{CASATI198777}%
  \BibitemOpen
  \bibfield  {author} {\bibinfo {author} {\bibfnamefont {G.}~\bibnamefont
  {Casati}}, \bibinfo {author} {\bibfnamefont {B.~V.}\ \bibnamefont
  {Chirikov}}, \bibinfo {author} {\bibfnamefont {D.~L.}\ \bibnamefont
  {Shepelyansky}},\ and\ \bibinfo {author} {\bibfnamefont {I.}~\bibnamefont
  {Guarneri}},\ }\bibfield  {title} {\bibinfo {title} {Relevance of classical
  chaos in quantum mechanics: The hydrogen atom in a monochromatic field},\
  }\href {https://doi.org/https://doi.org/10.1016/0370-1573(87)90009-3}
  {\bibfield  {journal} {\bibinfo  {journal} {Physics Reports}\ }\textbf
  {\bibinfo {volume} {154}},\ \bibinfo {pages} {77} (\bibinfo {year}
  {1987})}\BibitemShut {NoStop}%
\bibitem [{\citenamefont {Sagdeev}\ \emph {et~al.}(1988)\citenamefont
  {Sagdeev}, \citenamefont {Usikov},\ and\ \citenamefont
  {Zaslavsky}}]{sagdeev1988nonlinear}%
  \BibitemOpen
  \bibfield  {author} {\bibinfo {author} {\bibfnamefont {R.}~\bibnamefont
  {Sagdeev}}, \bibinfo {author} {\bibfnamefont {D.}~\bibnamefont {Usikov}},\
  and\ \bibinfo {author} {\bibfnamefont {G.}~\bibnamefont {Zaslavsky}},\ }\href
  {https://books.google.com.br/books?id=idRlyP37c3gC} {\emph {\bibinfo {title}
  {Nonlinear Physics: From the Pendulum to Turbulence and Chaos}}},\
  Contemporary concepts in physics\ (\bibinfo  {publisher} {Harwood Academic
  Publishers},\ \bibinfo {year} {1988})\BibitemShut {NoStop}%
\bibitem [{\citenamefont {Reichl}(2013)}]{reichl2013transition}%
  \BibitemOpen
  \bibfield  {author} {\bibinfo {author} {\bibfnamefont {L.}~\bibnamefont
  {Reichl}},\ }\href {https://books.google.com.br/books?id=slHuBwAAQBAJ} {\emph
  {\bibinfo {title} {The Transition to Chaos: Conservative Classical Systems
  and Quantum Manifestations}}},\ Institute for Nonlinear Science\ (\bibinfo
  {publisher} {Springer New York},\ \bibinfo {year} {2013})\BibitemShut
  {NoStop}%
\bibitem [{\citenamefont {Lichtenberg}\ and\ \citenamefont
  {Lieberman}(1983)}]{lieberman}%
  \BibitemOpen
  \bibfield  {author} {\bibinfo {author} {\bibfnamefont {A.~J.}\ \bibnamefont
  {Lichtenberg}}\ and\ \bibinfo {author} {\bibfnamefont {M.~A.}\ \bibnamefont
  {Lieberman}},\ }\href@noop {} {\emph {\bibinfo {title} {Regular and
  Stochastic Motion}}}\ (\bibinfo  {publisher} {Spring-Verlag},\ \bibinfo
  {year} {1983})\ \bibinfo {note} {applied Mathematics Sciences volume
  38}\BibitemShut {NoStop}%
\bibitem [{\citenamefont {de~Lima}\ \emph {et~al.}(2020)\citenamefont
  {de~Lima}, \citenamefont {de~Carvalho},\ and\ \citenamefont
  {Forlevesi}}]{PhysRevE.101.022207}%
  \BibitemOpen
  \bibfield  {author} {\bibinfo {author} {\bibfnamefont {E.~F.}\ \bibnamefont
  {de~Lima}}, \bibinfo {author} {\bibfnamefont {R.~E.}\ \bibnamefont
  {de~Carvalho}},\ and\ \bibinfo {author} {\bibfnamefont {M.~D.}\ \bibnamefont
  {Forlevesi}},\ }\bibfield  {title} {\bibinfo {title} {Nonchaotic laser pulse
  dissociation through deformed tori},\ }\href
  {https://doi.org/10.1103/PhysRevE.101.022207} {\bibfield  {journal} {\bibinfo
   {journal} {Phys. Rev. E}\ }\textbf {\bibinfo {volume} {101}},\ \bibinfo
  {pages} {022207} (\bibinfo {year} {2020})}\BibitemShut {NoStop}%
\bibitem [{\citenamefont {Leopold}\ and\ \citenamefont
  {Richards}(1985)}]{JGLeopold_1985}%
  \BibitemOpen
  \bibfield  {author} {\bibinfo {author} {\bibfnamefont {J.~G.}\ \bibnamefont
  {Leopold}}\ and\ \bibinfo {author} {\bibfnamefont {D.}~\bibnamefont
  {Richards}},\ }\bibfield  {title} {\bibinfo {title} {The effect of a resonant
  electric field on a one-dimensional classical hydrogen atom},\ }\href
  {https://doi.org/10.1088/0022-3700/18/16/021} {\bibfield  {journal} {\bibinfo
   {journal} {Journal of Physics B: Atomic and Molecular Physics}\ }\textbf
  {\bibinfo {volume} {18}},\ \bibinfo {pages} {3369} (\bibinfo {year}
  {1985})}\BibitemShut {NoStop}%
\bibitem [{\citenamefont {Burke}\ and\ \citenamefont
  {Mitchell}(2009)}]{PhysRevA.80.033416}%
  \BibitemOpen
  \bibfield  {author} {\bibinfo {author} {\bibfnamefont {K.}~\bibnamefont
  {Burke}}\ and\ \bibinfo {author} {\bibfnamefont {K.~A.}\ \bibnamefont
  {Mitchell}},\ }\bibfield  {title} {\bibinfo {title} {Chaotic ionization of a
  rydberg atom subjected to alternating kicks: Role of phase-space
  turnstiles},\ }\href {https://doi.org/10.1103/PhysRevA.80.033416} {\bibfield
  {journal} {\bibinfo  {journal} {Phys. Rev. A}\ }\textbf {\bibinfo {volume}
  {80}},\ \bibinfo {pages} {033416} (\bibinfo {year} {2009})}\BibitemShut
  {NoStop}%
\bibitem [{\citenamefont {Botina}\ \emph {et~al.}(1995)\citenamefont {Botina},
  \citenamefont {Rabitz},\ and\ \citenamefont {Rahman}}]{PhysRevA.51.923}%
  \BibitemOpen
  \bibfield  {author} {\bibinfo {author} {\bibfnamefont {J.}~\bibnamefont
  {Botina}}, \bibinfo {author} {\bibfnamefont {H.}~\bibnamefont {Rabitz}},\
  and\ \bibinfo {author} {\bibfnamefont {N.}~\bibnamefont {Rahman}},\
  }\bibfield  {title} {\bibinfo {title} {Optimal control of chaotic hamiltonian
  dynamics},\ }\href {https://doi.org/10.1103/PhysRevA.51.923} {\bibfield
  {journal} {\bibinfo  {journal} {Phys. Rev. A}\ }\textbf {\bibinfo {volume}
  {51}},\ \bibinfo {pages} {923} (\bibinfo {year} {1995})}\BibitemShut
  {NoStop}%
\bibitem [{\citenamefont {Joe-Wong}\ \emph {et~al.}(2013)\citenamefont
  {Joe-Wong}, \citenamefont {Ho}, \citenamefont {Long}, \citenamefont
  {Rabitz},\ and\ \citenamefont {Wu}}]{10.1063/1.4797498}%
  \BibitemOpen
  \bibfield  {author} {\bibinfo {author} {\bibfnamefont {C.}~\bibnamefont
  {Joe-Wong}}, \bibinfo {author} {\bibfnamefont {T.-S.}\ \bibnamefont {Ho}},
  \bibinfo {author} {\bibfnamefont {R.}~\bibnamefont {Long}}, \bibinfo {author}
  {\bibfnamefont {H.}~\bibnamefont {Rabitz}},\ and\ \bibinfo {author}
  {\bibfnamefont {R.}~\bibnamefont {Wu}},\ }\bibfield  {title} {\bibinfo
  {title} {{Topology of classical molecular optimal control landscapes in phase
  space}},\ }\href {https://doi.org/10.1063/1.4797498} {\bibfield  {journal}
  {\bibinfo  {journal} {The Journal of Chemical Physics}\ }\textbf {\bibinfo
  {volume} {138}},\ \bibinfo {pages} {124114} (\bibinfo {year}
  {2013})}\BibitemShut {NoStop}%
\bibitem [{\citenamefont {Rabitz}\ \emph {et~al.}(2023)\citenamefont {Rabitz},
  \citenamefont {Russell},\ and\ \citenamefont {Ho}}]{Rabitz2023}%
  \BibitemOpen
  \bibfield  {author} {\bibinfo {author} {\bibfnamefont {H.}~\bibnamefont
  {Rabitz}}, \bibinfo {author} {\bibfnamefont {B.}~\bibnamefont {Russell}},\
  and\ \bibinfo {author} {\bibfnamefont {T.-S.}\ \bibnamefont {Ho}},\
  }\bibfield  {title} {\bibinfo {title} {The surprising ease of finding optimal
  solutions for controlling nonlinear phenomena in quantum and classical
  complex systems},\ }\href {https://doi.org/10.1021/acs.jpca.3c01896}
  {\bibfield  {journal} {\bibinfo  {journal} {The Journal of Physical Chemistry
  A}\ }\textbf {\bibinfo {volume} {127}},\ \bibinfo {pages} {4224} (\bibinfo
  {year} {2023})}\BibitemShut {NoStop}%
\bibitem [{\citenamefont {Constantoudis}\ and\ \citenamefont
  {Nicolaides}(2001)}]{nicolaides}%
  \BibitemOpen
  \bibfield  {author} {\bibinfo {author} {\bibfnamefont {V.}~\bibnamefont
  {Constantoudis}}\ and\ \bibinfo {author} {\bibfnamefont {C.~A.}\ \bibnamefont
  {Nicolaides}},\ }\bibfield  {title} {\bibinfo {title} {Nonhyperbolic escape
  and changes in phase-space stability structures in laser-induced multiphoton
  dissociation of a diatomic molecule},\ }\href@noop {} {\bibfield  {journal}
  {\bibinfo  {journal} {Phys. Rev. E}\ }\textbf {\bibinfo {volume} {64}},\
  \bibinfo {pages} {056211} (\bibinfo {year} {2001})}\BibitemShut {NoStop}%
\bibitem [{\citenamefont {Iñarrea}\ \emph {et~al.}(2019)\citenamefont
  {Iñarrea}, \citenamefont {Lanchares}, \citenamefont {Palacián},
  \citenamefont {Pascual}, \citenamefont {Salas},\ and\ \citenamefont
  {Yanguas}}]{INARREA201994}%
  \BibitemOpen
  \bibfield  {author} {\bibinfo {author} {\bibfnamefont {M.}~\bibnamefont
  {Iñarrea}}, \bibinfo {author} {\bibfnamefont {V.}~\bibnamefont {Lanchares}},
  \bibinfo {author} {\bibfnamefont {J.~F.}\ \bibnamefont {Palacián}}, \bibinfo
  {author} {\bibfnamefont {A.~I.}\ \bibnamefont {Pascual}}, \bibinfo {author}
  {\bibfnamefont {J.~P.}\ \bibnamefont {Salas}},\ and\ \bibinfo {author}
  {\bibfnamefont {P.}~\bibnamefont {Yanguas}},\ }\bibfield  {title} {\bibinfo
  {title} {Effects of a soft-core coulomb potential on the dynamics of a
  hydrogen atom near a metal surface},\ }\href
  {https://doi.org/https://doi.org/10.1016/j.cnsns.2018.07.039} {\bibfield
  {journal} {\bibinfo  {journal} {Communications in Nonlinear Science and
  Numerical Simulation}\ }\textbf {\bibinfo {volume} {68}},\ \bibinfo {pages}
  {94} (\bibinfo {year} {2019})}\BibitemShut {NoStop}%
\bibitem [{\citenamefont {Gebremedhin}\ and\ \citenamefont
  {Weatherford}(2014)}]{PhysRevE.89.053319}%
  \BibitemOpen
  \bibfield  {author} {\bibinfo {author} {\bibfnamefont {D.~H.}\ \bibnamefont
  {Gebremedhin}}\ and\ \bibinfo {author} {\bibfnamefont {C.~A.}\ \bibnamefont
  {Weatherford}},\ }\bibfield  {title} {\bibinfo {title} {Calculations for the
  one-dimensional soft coulomb problem and the hard coulomb limit},\ }\href
  {https://doi.org/10.1103/PhysRevE.89.053319} {\bibfield  {journal} {\bibinfo
  {journal} {Phys. Rev. E}\ }\textbf {\bibinfo {volume} {89}},\ \bibinfo
  {pages} {053319} (\bibinfo {year} {2014})}\BibitemShut {NoStop}%
\bibitem [{\citenamefont {Javanainen}\ \emph {et~al.}(1988)\citenamefont
  {Javanainen}, \citenamefont {Eberly},\ and\ \citenamefont
  {Su}}]{PhysRevA.38.3430}%
  \BibitemOpen
  \bibfield  {author} {\bibinfo {author} {\bibfnamefont {J.}~\bibnamefont
  {Javanainen}}, \bibinfo {author} {\bibfnamefont {J.~H.}\ \bibnamefont
  {Eberly}},\ and\ \bibinfo {author} {\bibfnamefont {Q.}~\bibnamefont {Su}},\
  }\bibfield  {title} {\bibinfo {title} {Numerical simulations of multiphoton
  ionization and above-threshold electron spectra},\ }\href
  {https://doi.org/10.1103/PhysRevA.38.3430} {\bibfield  {journal} {\bibinfo
  {journal} {Phys. Rev. A}\ }\textbf {\bibinfo {volume} {38}},\ \bibinfo
  {pages} {3430} (\bibinfo {year} {1988})}\BibitemShut {NoStop}%
\bibitem [{\citenamefont {Su}\ \emph {et~al.}(1990)\citenamefont {Su},
  \citenamefont {Eberly},\ and\ \citenamefont
  {Javanainen}}]{PhysRevLett.64.862}%
  \BibitemOpen
  \bibfield  {author} {\bibinfo {author} {\bibfnamefont {Q.}~\bibnamefont
  {Su}}, \bibinfo {author} {\bibfnamefont {J.~H.}\ \bibnamefont {Eberly}},\
  and\ \bibinfo {author} {\bibfnamefont {J.}~\bibnamefont {Javanainen}},\
  }\bibfield  {title} {\bibinfo {title} {Dynamics of atomic ionization
  suppression and electron localization in an intense high-frequency radiation
  field},\ }\href {https://doi.org/10.1103/PhysRevLett.64.862} {\bibfield
  {journal} {\bibinfo  {journal} {Phys. Rev. Lett.}\ }\textbf {\bibinfo
  {volume} {64}},\ \bibinfo {pages} {862} (\bibinfo {year} {1990})}\BibitemShut
  {NoStop}%
\bibitem [{\citenamefont {Li}\ and\ \citenamefont
  {Jones}(2014)}]{PhysRevLett.112.143006}%
  \BibitemOpen
  \bibfield  {author} {\bibinfo {author} {\bibfnamefont {S.}~\bibnamefont
  {Li}}\ and\ \bibinfo {author} {\bibfnamefont {R.~R.}\ \bibnamefont {Jones}},\
  }\bibfield  {title} {\bibinfo {title} {Ionization of excited atoms by intense
  single-cycle thz pulses},\ }\href
  {https://doi.org/10.1103/PhysRevLett.112.143006} {\bibfield  {journal}
  {\bibinfo  {journal} {Phys. Rev. Lett.}\ }\textbf {\bibinfo {volume} {112}},\
  \bibinfo {pages} {143006} (\bibinfo {year} {2014})}\BibitemShut {NoStop}%
\bibitem [{\citenamefont {Bestle}\ \emph {et~al.}(1993)\citenamefont {Bestle},
  \citenamefont {Akulin},\ and\ \citenamefont {Schleich}}]{PhysRevA.48.746}%
  \BibitemOpen
  \bibfield  {author} {\bibinfo {author} {\bibfnamefont {J.}~\bibnamefont
  {Bestle}}, \bibinfo {author} {\bibfnamefont {V.~M.}\ \bibnamefont {Akulin}},\
  and\ \bibinfo {author} {\bibfnamefont {W.~P.}\ \bibnamefont {Schleich}},\
  }\bibfield  {title} {\bibinfo {title} {Classical and quantum stabilization of
  atoms in intense laser fields},\ }\href
  {https://doi.org/10.1103/PhysRevA.48.746} {\bibfield  {journal} {\bibinfo
  {journal} {Phys. Rev. A}\ }\textbf {\bibinfo {volume} {48}},\ \bibinfo
  {pages} {746} (\bibinfo {year} {1993})}\BibitemShut {NoStop}%
\bibitem [{\citenamefont {Krempl}\ \emph {et~al.}(1992)\citenamefont {Krempl},
  \citenamefont {Eisenhammer}, \citenamefont {H\"ubler}, \citenamefont
  {Mayer-Kress},\ and\ \citenamefont {Milonni}}]{PhysRevLett.69.430}%
  \BibitemOpen
  \bibfield  {author} {\bibinfo {author} {\bibfnamefont {S.}~\bibnamefont
  {Krempl}}, \bibinfo {author} {\bibfnamefont {T.}~\bibnamefont {Eisenhammer}},
  \bibinfo {author} {\bibfnamefont {A.}~\bibnamefont {H\"ubler}}, \bibinfo
  {author} {\bibfnamefont {G.}~\bibnamefont {Mayer-Kress}},\ and\ \bibinfo
  {author} {\bibfnamefont {P.~W.}\ \bibnamefont {Milonni}},\ }\bibfield
  {title} {\bibinfo {title} {Optimal stimulation of a conservative nonlinear
  oscillator: Classical and quantum-mechanical calculations},\ }\href
  {https://doi.org/10.1103/PhysRevLett.69.430} {\bibfield  {journal} {\bibinfo
  {journal} {Phys. Rev. Lett.}\ }\textbf {\bibinfo {volume} {69}},\ \bibinfo
  {pages} {430} (\bibinfo {year} {1992})}\BibitemShut {NoStop}%
\bibitem [{\citenamefont {Kirk}(1970)}]{kirk}%
  \BibitemOpen
  \bibfield  {author} {\bibinfo {author} {\bibfnamefont {D.~E.}\ \bibnamefont
  {Kirk}},\ }\href@noop {} {\emph {\bibinfo {title} {Optimal Control Theory An
  Introduction}}}\ (\bibinfo  {publisher} {Prentice-Hall},\ \bibinfo {year}
  {1970})\BibitemShut {NoStop}%
\end{thebibliography}

%

\end{document}